\documentstyle[preprint,aps]{revtex}
\def\vA{v_{\scriptscriptstyle A}}
\def\tA{t_{\scriptscriptstyle A}}
\def\tp{t_{\scriptscriptstyle p}}
\def\grad{\nabla_{\scriptscriptstyle\perp}}
\def\lapl{\grad\sp2}
\def\dt{\partial_t}
\def\dx{\partial_x}
\def\dy{\partial_y}
\def\dz{\partial_z}
\def\zz{\hat{\bf z}}
\def\l0{l_p}
\def\Eqn#1{equation (\ref{eq:#1})}
\def\Eqs#1#2{equations (\ref{eq:#1})-(\ref{eq:#2})}
\def\Fig#1{Figure~\ref{fig:#1}}

\def\apj{{\it Astrophys. J.}}

\begin{document}

\title{Energy spectrum of turbulent fluctuations in boundary
driven reduced magnetohydrodynamics}

\author{Pablo Dmitruk$^1$, Daniel O. G\'omez$^2$ and William H. Matthaeus$^1$}

\address{
$^1$Bartol Research Institute, University of Delaware,
       Newark, DE 19716 \\
$^2$ Departmento de F\'{\i}sica, Facultad de Ciencias Exactas y Naturales,
     Universidad de Buenos Aires, Ciudad Universitaria,
     1428 Buenos Aires, Argentina}

\maketitle

\begin{abstract}
The nonlinear dynamics of a bundle of magnetic flux ropes driven by 
stationary fluid motions at their endpoints is studied, by performing numerical simulations 
of the magnetohydrodynamic (MHD) equations. The development of MHD 
turbulence is shown, where the system reaches a state that is characterized by the ratio 
between the Alfv\'en time (the time for incompressible MHD waves to travel along 
the field lines) and the convective time scale of the driving motions. 
This ratio of time scales determines the energy spectra and the relaxation toward different 
regimes ranging from weak to strong turbulence. A connection is made with 
phenomenological theories for the energy spectra in MHD turbulence.
\end{abstract}

\section{Introduction}

The search for universal scaling properties in magnetohydrodynamic turbulence,
of the type developed by Kolmogorov \cite{Kolmogorov41} for 
hydrodynamic turbulence,
has been the subject of many theoretical and numerical studies, 
since the pioneering phenomenological arguments put forward independently by 
Iroshnikov and Kraichnan
\cite{Iroshnikov,Kraichnan65} (hereafter IK). The phenomenological arguments of Kolomogorov, which
give rise to the famous power law $E_k \sim k^{-5/3}$ for 
the omnidirectional energy spectrum $E_k$ in terms of the wavenumber $k$,
were modified by IK to include magnetic field effects, 
deriving an energy spectrum power law like $E_k \sim k^{-3/2}$. 
An alternative point of view, however,
was proposed \cite{Fyfe77} indicating that the original Kolmogorov
scenario is still applicable in certain cases to MHD as well as to
hydrodynamics. 
Along similar lines, a phenomenological theory was proposed 
in Ref. \cite{MatthaeusZhou89}
for a steady inertial range spectral law that
reduces to the IK and Kolmogorov laws in appropriate limits.
To distinguish between the two power laws is a difficult task
both observationally and numerically \cite{MullerBiskamp00},
partly because the power law indexes are
too close, but also because assumptions on the theories, such as homogeneity,
isotropy and time stationarity are often lost. 

For instance, when a strong, externally supported, magnetic field is present,
the key assumption of isotropy breaks down. It has been shown 
\cite{Shebalin83,Oughton94,GoldreichSridhar97,KinneyMcWilliams98} 
that the cascade renders itself anisotropic, and that the spectrum
in the direction of the magnetic field becomes strongly suppressed. 
This situation also
holds for compressible MHD, where a variance anisotropy
is also found \cite{Matthaeus96}, 
i.e., components of the fluctuating fields in the strong
magnetic field direction are small compared to the transverse components.
On theoretical grounds, this configuration has been found 
\cite{ZankMatthaeus92} to be appropriately described, 
in the low-frequency limit, 
by the so called Reduced MHD \cite{Strauss76,Montgomery82} approximation (RMHD). 

Phenomenological theories alternative to the IK
scenario \cite{Iroshnikov,Kraichnan65} when a strong magnetic field is present
have been also presented to describe the perpendicular spectrum in anisotropic
conditions \cite{NgBhattacharjee97,GoldreichSridhar97}. 
Recently, a formal closure model has been introduced, 
which obtains a kinetic equation for the anisotropic energy spectrum 
in the limit known as {\it weak MHD turbulence} \cite{Galtier00}.

Besides the anisotropy introduced by a strong magnetic field,
another effect that may modify the cascade situation is the presence
of driving boundaries. We intend to address the effect of both 
a strong external magnetic field and the driving boundaries on MHD turbulence, 
by considering the particular case of a plasma under a 
strong uniform field in the z-direction, limited within the 
transverse planes $z=0$ and $z=L_z$ where an imposed velocity field is 
applied.
This case is inspired in the theoretical 
model of a coronal magnetic loop, which is forced
through convective motions at its footpoints. Such a model loop, proposed
in Ref. \cite{Parker72}, has been widely studied in many different contexts,
although not necessarily in a turbulent scenario.
Most studies address the issue of current sheet formation
\cite{VanBallegooijen86,Longcope94,Mikic89}, non-steady 
reconnection \cite{HendrixVanHoven96,Milano99} and 
coronal heating
\cite{GomezFerro92,GalsgaardNordlund96,HendrixVanHoven96,DmitrukGomez99,Mandrini00}.
However, this paper is not aimed at studying the dynamics of coronal magnetic loops.
Instead, we focus our analysis on the (steady) perpendicular energy spectrum 
of boundary driven RMHD
and make a connection with the phenomenological theories of MHD turbulence.
We pay particular attention to the issues of numerical resolution required
to yield well resolved turbulent spectra from direct numerical simulations. 

The organization of the paper is as follows: in section II we 
present our model and the RMHD equations. Section III contains a description of
the boundary conditions and the numerical code. Section IV describes
the statistically steady regime and the dissipation rate scaling law.
In Section V we show the numerically obtained energy spectra and
discuss the connection with phenomenological theories of turbulence.
Issues of numerical resolution are also addressed.
Section VI discusses the typical dissipative structures. Section VII
contains the conclusions.

\section{Model and equations}

\begin{figure}
   \caption[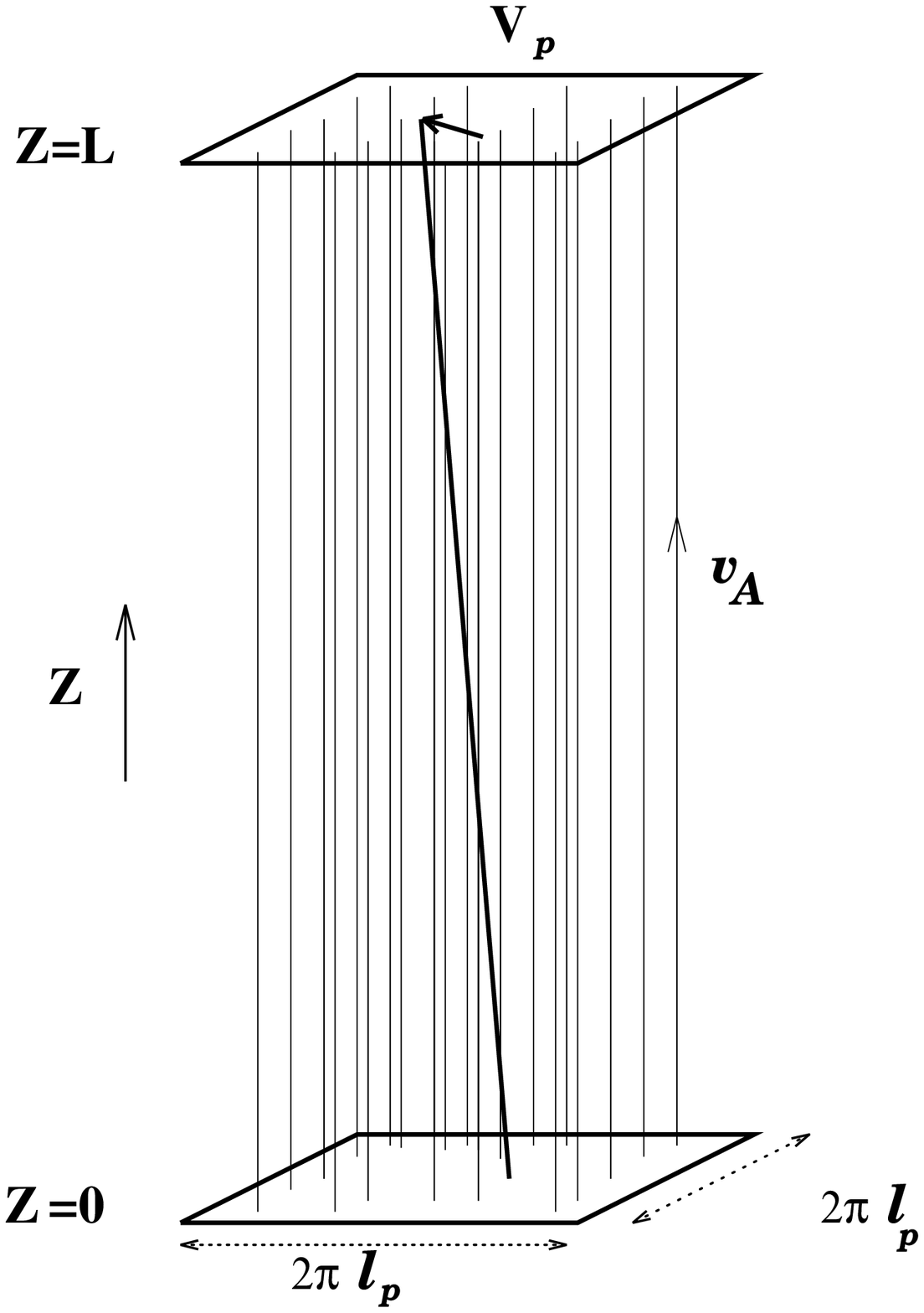]{Cartoon of the reduced MHD setup.
   }
   \label{fig:1}
\end{figure}

Let us assume a low-$\beta$ magnetofluid ($\beta$ is the ratio 
of gas pressure to magnetic pressure) permeated by an initially 
uniform magnetic field ${\bf B} = B_0 \zz$, which is 
elongated along the $\zz$-direction (i.e. $L_z \gg L_\perp $), 
as shown in \Fig{1}. 
Under these conditions, the so-called reduced magnetohydrodynamic 
equations (RMHD) are applicable 
\cite{Strauss76,Montgomery82,ZankMatthaeus92}
to describe the dominant low-frequency non-linear motions. 
Within this approximation, the magnetic and velocity 
fields are both divergence-free and can be expressed as
\begin{eqnarray}
{\bf B} = B_0\zz\ +\  \grad\times (a\zz) \\
{\bf u} = \grad\times (\psi\zz)
\label{eq:ByU}
\end{eqnarray}
where $a (x, y, z, t)$ is the magnetic flux function 
and $\psi (x, y, z, t)$ is the stream 
function. The RMHD equations in terms of these scalar potentials are:
\begin{equation}
   \dt a  = \vA \dz \psi + [ \psi , a ] + \eta \lapl a
\label{eq:dta}
\end{equation}
\begin{equation}
   \dt w = \vA \dz j + [ \psi , w ] - [ a , j ] + \nu \lapl w
\label{eq:dtw}
\end{equation}
where $w (x, y, z, t) = - \lapl\psi $ is the parallel vorticity 
and $j (x, y, z, t) = - \lapl a $ is the parallel current density. 
The coefficient $\vA = B_0/\sqrt{4\pi\rho}$ 
is the Alfv\'en speed, $\nu$ is the kinematic viscosity 
and $\eta $ is the magnetic diffusivity.  
The non-linear terms in these equations are expressed in terms of standard 
Poisson brackets, i.e. $[u,v]=(\dx u)(\dy v) - (\dy u)(\dx v)$.

This particular theoretical setup is relevant to several plasma applications, 
ranging from tokamaks to magnetic loops in the solar corona. The RMHD framework 
has been used to study the dynamics of coronal loops in solar active regions 
\cite{GomezFerro92,Longcope94,HendrixVanHoven96,DmitrukGomez99}. The axial (and 
approximately constant) magnetic field in these loops has both ends or footpoints 
anchored in the solar photosphere. The photosphere is a high-$\beta$ plasma, which 
is also convectively turbulent. These convective motions at the photosphere, in 
turn move the magnetic fieldlines around, and drive the coronal part of the loops 
into a rather complex dynamical scenario.

We specify the velocity fields at the boundaries as
\begin{equation}
  \psi (z = 0) = 0,\ \ \ \ \ \ \psi (z = L_z) = \Psi (x,y)
\label{eq:psiphot}
\end{equation}
where $\Psi (x,y)$ is the stream function which describes stationary and 
incompressible footpoint motions.  The strength of this external velocity field 
is proportional to a typical velocity $U_p$.

To transform \Eqs{dta}{dtw} into their dimensionless form, 
we choose $\l0 = L_\perp /(2 \pi ) $ 
and $L_z$ as the units for transverse and longitudinal distances. 
Since the dimensions of all 
physical quantities involved in these equations can be expressed 
as combinations of {\it length} and 
{\it time}, let us choose  $\tA\equiv L_z/\vA$ as the time unit. 
The dimensionless RMHD equations are:
\begin{equation}
   \dt a  = \dz \psi + [ \psi , a ] + {1 \over S} \lapl a
\label{eq:dta2}
\end{equation}

\begin{equation}
   \dt w = \dz j + [ \psi , w ] - [ a , j ] + {1 \over R} \lapl w
\label{eq:dtw2}
\end{equation}

\noindent
where $S^{-1}= {{\eta\tA}\over\l0^2}$ and $R^{-1}= {{\nu\tA}\over\l0^2}$ are
the (dimensionless) magnetic and kinetic dissipation coefficients. Hereafter, we
will consider the case $S = R$, and thus the (common) dissipation coefficient will be
the only dimensionless parameter explicitly present in \Eqs{dta2}{dtw2} .

\section{Description of the code and boundary driving}

We numerically integrated \Eqs{dta2}{dtw2}. 
To this end, $\psi$ and $a$ are expanded in 
Fourier modes in each $(x,y)$ plane ($0 \le x,y \le 2\pi$ and $0\le z \le 1$). 
The corresponding 
Fourier coefficients $\psi_{\bf k} (z,t)$ and $a_{\bf k} (z,t)$ 
are evolved in time using a 
semi-implicit scheme: linear terms are treated in a fully implicit fashion, 
while nonlinear terms
are evolved using a second order Runge-Kutta scheme. 
Also, nonlinear terms are evaluated following 
a $2/3$ fully dealiased (see Ref. \cite{Canuto88}) pseudo-spectral technique. 

To compute $z$-derivatives we use a standard method of finite differences 
in a staggered 
regular grid (see for instance Refs. \cite{Strauss76,Longcope94}) of $N_z+1$ points. 
The stream function is 
computed on points $z_i=i/N_z ~(i=0,\dots,N_z)$, 
while the magnetic flux function is computed 
on $z_{i+1/2}=(i+1/2)/N_z ~(i=0,\dots,N_z-1)$. Boundary conditions for the 
stream function $\psi$ are given at the plates $z=0$ and $z=1$. 
Therefore, \Eqn{dtw2} is not 
integrated on these planes, but it is evolved in time in all the 
internal gridpoints $z_i=i/N_z ~(i=1,\dots,N_z-1)$.

We specify the stream function in \Eqn{psiphot} as
\begin{eqnarray}
\Psi_{\bf k} & = & \Psi_0 =\tA / \tp,\ \ \ \ \ \ {\rm if} ~~3 < k~\l0 < 4\\ \nonumber
\Psi_{\bf k} & = & 0\ \ \ \ \ \ \ \ \ \ \ \ \ \ \ \ \ \ \ \ \ \ \  {\rm elsewhere}
\label{eq:psi0}
\end{eqnarray}

This expression imitates a stationary pattern of eddy motions of diameters between 
$L_\perp/4$ and $L_\perp/3$, rotation speeds $U_p$ and typical turnover  times 
$\tp = \l0 /U_p$. Our choice of this narrowband and non-random forcing ensures that the 
broadband energy spectra that we obtain are exclusively determined by the nonlinear 
nature of the MHD equations. 

The strength of the external driver is quantified by the 
dimensionless factor 
$\Psi_0 =\tA / \tp$ in \Eqn{psi0}, which is given by the ratio of the 
Alfven time of the system (i.e. the response time to an impulse applied at 
the boundary) to the timescale of the driver itself 
(i.e. the eddy turnover time).
Note that the velocity fields that we are imposing at the boundaries
are stationary.
For non-stationary footpoint motions, which might for instance 
represent wave activity 
or time-correlated random flows, other timescales should be considered in 
connection with the driver.

\section{Stationary regimes and scaling law}

As mentioned in Section 2 , the dissipation coefficient $S$ (with $S = R $) is the only 
dimensionless parameter present in \Eqs{dta2}{dtw2}. Just as important is the 
dimensionless factor $\tA/\tp$ introduced by the external force applied at the 
boundary. Therefore, we are left with these two dimensionless numbers to 
characterize the solutions of the RMHD equations [i.e. \Eqs{dta2}{dtw2} with the 
boundary condition given by \Eqn{psiphot} and \Eqn{psi0}].

From purely dimensional considerations, we know that for any physical quantity, 
its dimensionless version $Q$ should be an arbitrary function of the only two 
dimensionless parameters of the problem, i.e.
\begin{equation}
  Q = {\cal F} ( Q_1 , Q_2 ),\ \ \ \ Q_1 = {\tA\over\tp},\ \ \ Q_2 = S
\label{eq:Q}
\end{equation}

For instance, for the dissipation rate per unit mass $\epsilon$
\begin{equation}
  Q = \epsilon {\tA^3\over\l0^2} = {\cal F} ( {\tA\over\tp} , S )
\label{eq:Qeps1}
\end{equation}

A sequence of numerical simulations for different values of the dissipation 
coefficient performed in Ref. \cite{DmitrukGomez99} (see also 
Ref. \cite{Hendrix96}), indicated that the dependence of the dissipation rate 
with S is rather mild. This relative insensitivity of the dissipation rate 
with the dissipation coefficient is consistent with similar results obtained 
from experiments in purely hydrodynamic turbulence \cite{Sreeni,Batchelor,Frisch96}.
It is also consistent with one of Kolmogorov's hypothesis for statistically
steady turbulent
regimes at very large Reynolds numbers, which assumes that
 the dissipation rate remains finite in the limit of vanishing viscosity 
\cite{Frisch96}. Therefore, if we assume that the dissipation rate in 
statistically steady
turbulent regimes in RMHD is approximately independent of the dissipation coefficient $S$, 
we readily obtain that the dissipation rate will only depend on the time ratio $\tA/\tp$. 
A second series of simulations \cite{DmitrukGomez99}, in this case for different 
values of the ratio $\tA/\tp$, led to the following  expression for the function 
${\cal F}$, 

\begin{equation}
   \epsilon = {\l0^2\over\tA^3} 
\Big( {\tA\over\tp} \Big)^s, \ \ \ \ \ \ \ \ \ \ \ \ \ \ \ s = 1.51 \pm 0.04
\label{eq:Qeps3}
\end{equation}

\begin{figure}
   \caption[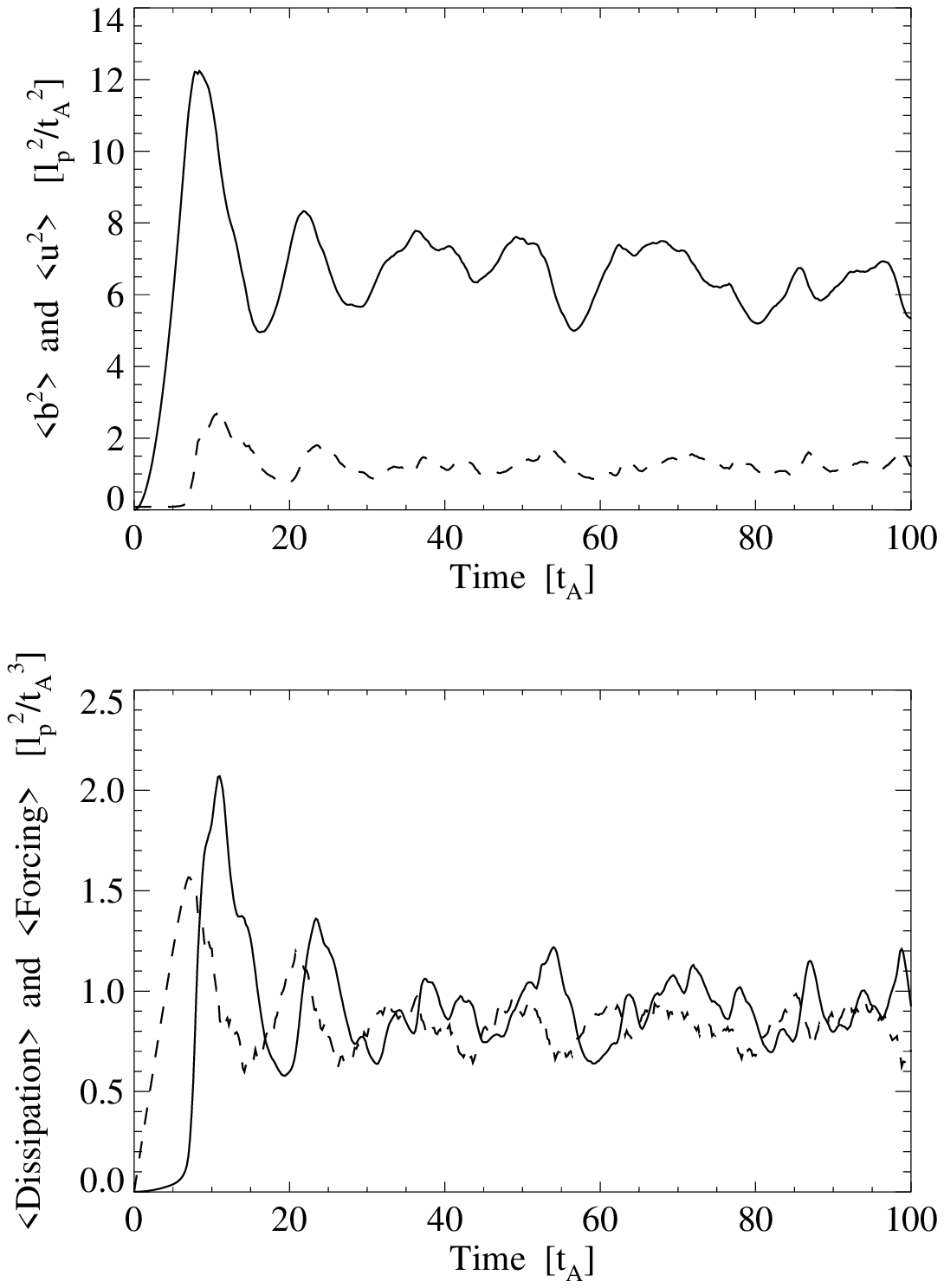]{
   Top: magnetic and kinetic energy (dashed) vs time from a run with
   $S=2000$, $t_A/t_p=0.5$. Bottom: Dissipation rate and forcing (dashed)
   for the same run.
   }
   \label{fig:2}
\end{figure}

In the present paper, we performed numerical simulations of \Eqs{dta2}{dtw2} with
$S = 2000$ and different values of the ratio $\tA /\tp$ in the
range $[ 0.1 , 1]$. Note that $\tA /\tp$ can also be written as 
$\tA /\tp = (L_z/\l0)/(\vA/U_p)$. For consistency with the RMHD approximation, both $L_z/\l0$ 
and $\vA/U_p$ should remain much larger than unity. However, as mentioned above, the only 
relevant parameter in this problem is $\tA /\tp$, which is free to take any value.
An additional simulation with $\tA/\tp=2$ and $S=800$ is also performed.
Numerical resolution ranges from $256\times256\times64$
to $2048\times 2048\times 128$ gridpoints.
A Beowulf PC cluster is employed to perform those runs, and a parallel
RMHD code has been designed. Since finite-differences
are used in the $z$-direction, while a pseudospectral method is
employed for the transverse directions, a very efficient parallelization
is achieved by performing the transverse gradients locally in each machine,
while employing communication between machines only to perform the
$z$-derivatives. The typical behavior of the magnetic and kinetic energy as a function 
of time is shown in the top panel of \Fig{2}, while the bottom panel shows the 
dissipation rate for the particular case of $\tA /\tp = 0.5$ with $256\times256\times64$
resolution. After an initial transient, the dissipation rate is seen 
to approach a statistically steady level.

\section{Energy spectra}

Energy spectra of turbulent fluctuations are essential both for 
phenomenological and statistical theories of MHD turbulence.
We compute the perpendicular spectra at each plane and integrate
in $z$ to obtain a {\it perpendicular spectrum},
\begin{equation}
E_{k_{\perp}} = \frac{1}{2} \int \sum_{k_{\perp} < k_x^2+k_y^2 <
                         k_{\perp} + \Delta k}
                         \lbrack ~|{\bf b}(k_x,k_y,z)|^2 +
                         |{\bf u}(k_x,k_y,z)|^2 ~\rbrack ~dz
\end{equation}
where the total energy can be obtained as $E=\int E_{k_{\perp}} dk_{\perp}$.

As we have pointed out in the previous section, the system reaches
an approximately steady state where the forcing is compensated by
dissipation (\Fig{2} bottom). The first stage of the system evolution is 
followed using a low resolution simulation. Once the steady state is
achieved, a higher resolution simulation is started, by taking
the Fourier coefficients from the previous run and padding the additional
high $k$ coefficients to zero. After a short transient, those additional
coefficients are populated by the nonlinear cascade in a few
eddy turnover times. In this way, a high
resolution steady state spectrum is obtained in a relatively shorter computer
time. The high resolution runs ($2048\times2048\times64$) are however
computationally demanding, even for runs of a few eddy turnover times.

\begin{figure}
   \caption[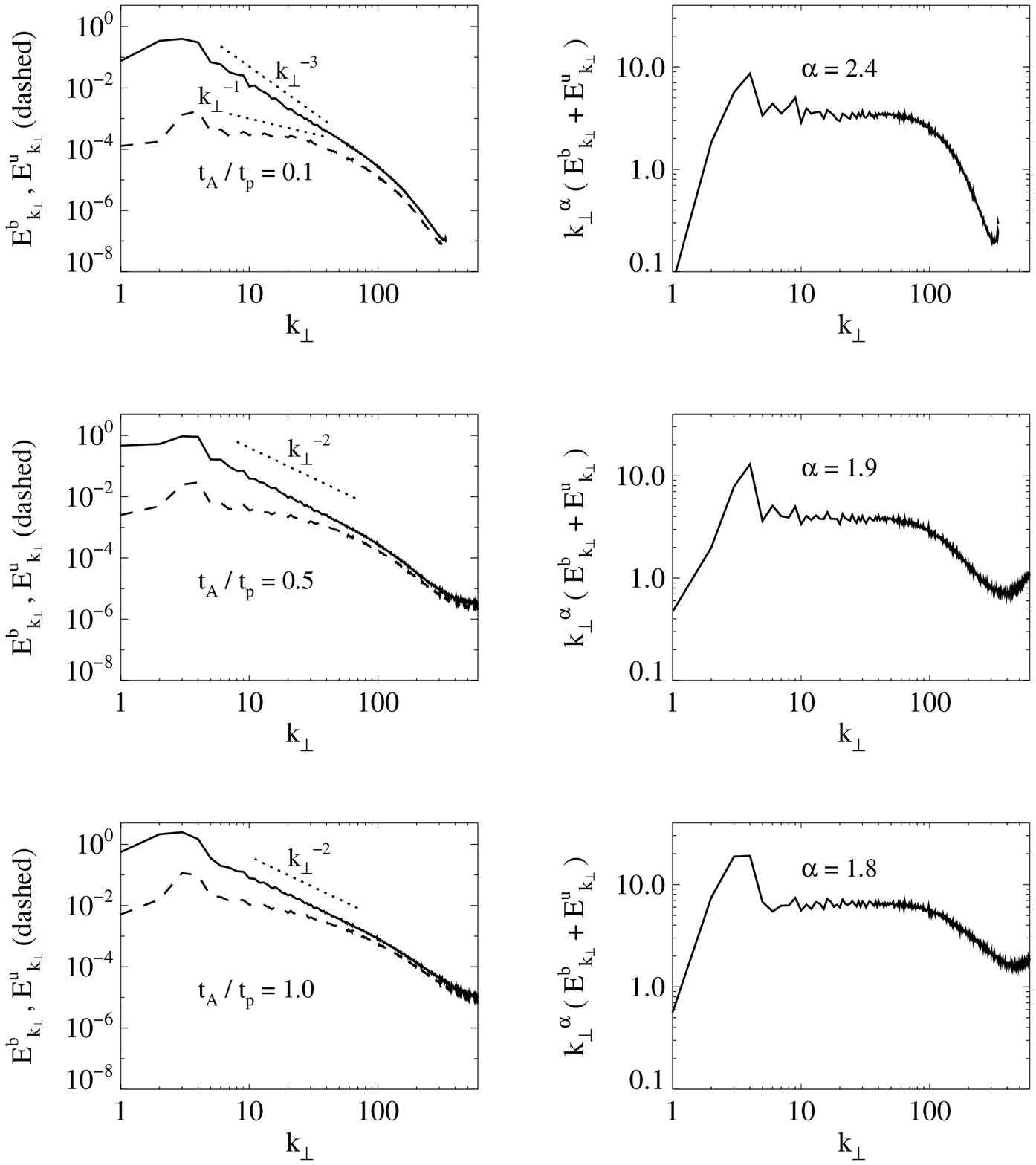]{
     Perpendicular energy spectra from $S=2000$ runs and different 
     values of the ratio $t_A/t_p=0.1,~0.5,~1$. 
     Resolution is $1024 \times 1024 \times 64,~2048 \times 2048 \times 64,
     ~2048 \times 2048 \times 128$ respectively.
     On the left panels, magnetic and
     kinetic energy (dashed) perpendicular spectra. 
     The dotted lines show exact power laws given for reference.
     On the right, total ``compensated''
     perpendicular spectra and corresponding power index value $\alpha$.
    }
   \label{fig:3}
\end{figure}

\begin{figure}
   \caption[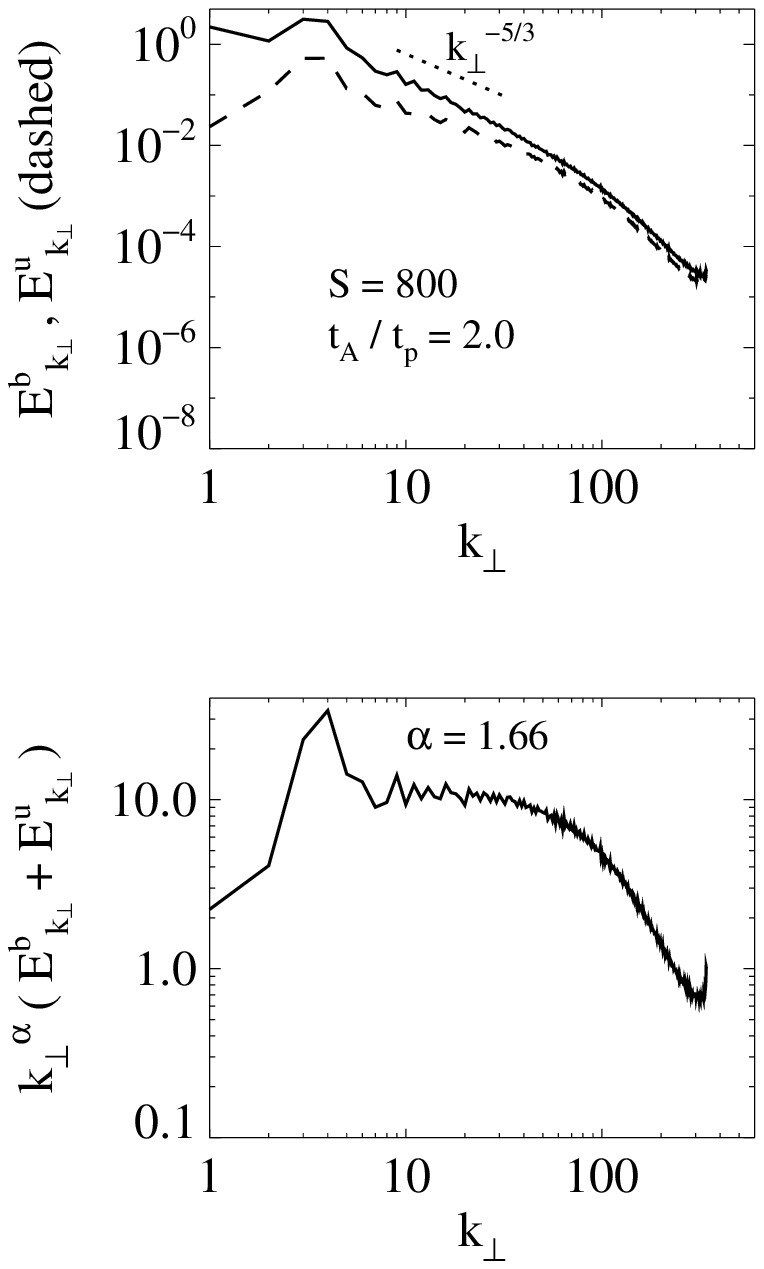]{Spectra and compensated spectra for a run with
   $S=800$, $t_A/t_p=2$.
    }
   \label{fig:4}
\end{figure}

In \Fig{3} we show the magnetic and kinetic energy perpendicular spectra 
for different values of the quantity $t_A/t_p$. The right panel
plots show the compensated spectra (magnetic plus kinetic), 
which allow determination of
the power law index $\alpha$ observed in the inertial range 
where $E_{k_{\perp}} \sim k_{\perp}^{-\alpha}$.
A first clear result observed in these plots is that the power
law index depends on the value of $t_A/t_p$. For low values
of $t_A/t_p$ a rather steep spectra ($\alpha > 2$) is
observed. At intermediate values of $t_A/t_p$ the power law index
is approximately equal to $\alpha = 2$. 
For larger values of $t_A/t_p$ the spectral 
law becomes flatter ($\alpha < 2$). As shown in \Fig{4}, 
when $t_A/t_p=2$ 
a value close to $\alpha = 5/3$ is obtained. For this last simulation 
a lower value of $S=800$
has to be employed, in order to have a well resolved spectrum (see
discussion below). 

Phenomenological theories 
\cite{NgBhattacharjee97,BhattacharjeeNg01}
and a more formal theory based on closures \cite{Galtier00} have
been proposed to explain a $k^{-2}$ spectra in a regime known
as the {\it weak turbulence} limit. 
An alternative phenomenological derivation of this power law can be
done using the general approach proposed in Ref. \cite{MatthaeusZhou89}
in terms of the triple correlation time. 
The energy supply $\epsilon$ is assumed to be equal to the $k$-independent
spectral transfer rate. As stated in Ref. \cite{MatthaeusZhou89}, if
$\tau_3(k)$ is the time scale for decay of triple correlations, which
induce the spectral transfer to wavenumbers higher than $k$,
then \cite{Kraichnan65,MatthaeusZhou89}
\begin{equation}
\epsilon \sim \tau_3(k) \frac{k E_k}{\tau_{nl}^2(k)}
\label{eq:phenom}
\end{equation}
where $\tau_{nl} (k) \sim (k u_k)^{-1}$ is the non-linear time.  
Equation (\ref{eq:phenom}) implies a spectral transfer time
$\tau_s \sim \tau_{nl}^2/\tau_3$.
If $\tau_3(k) \sim \tau_{nl}(k)$ and equipartition between magnetic
and kinetic energy is assumed $E^b_k \sim E^u_k \sim u_k^2/k$,
the usual $E_k \sim k^{-5/3}$ Kolmogorov scaling is recovered. 
When $\tau_3(k) \sim \tau_A(k) = (k \bar{b})^{-1}$, 
where $\bar{b}$ is the rms of magnetic fluctuations,
and isotropy and equipartition is assumed,
the IK scaling $E_k \sim k^{-3/2}$ is obtained.
If there is a strong external magnetic field the energy
cascade is anisotropic and
$\tau_{nl} \sim (k_{\perp} u_{k_{\perp}})^{-1}$. 
For this case, the appropriate
triple correlation time $\tau_3$ is the
Alfven wave crossing time (in the direction of the external field) 
$\tau_3 \sim t_A=L_z/V_A$,
a fixed external quantity independent of $k_{\perp}$
(this is different from the $k$-dependent Alfven time $\tau_A (k)$ 
associated with fluctuations in an isotropic situation).
Again assuming equipartition,
\begin{equation}
\epsilon \sim t_A k_{\perp} E_{k_{\perp}} 
                  k_{\perp}^2 u_{k_{\perp}}^2
         \sim t_A k_{\perp}^4 E_{k_{\perp}}^2
\label{eq:weakturb}
\end{equation}
and a $E_{k_{\perp}} \sim k^{-2}_{\perp}$ spectra is obtained. 
This is equivalent
to the wave packet interaction approach presented 
in Ref. \cite{NgBhattacharjee97},
based on the IK scenario \cite{Iroshnikov,Kraichnan65} but under 
strong anisotropic conditions.
It is also interesting that an essentially similar argument 
\cite{Zhou95} has been applied to the case of purely hydrodynamic
turbulence in the presence of strong rotation, where the 
triple correlation time is set equal to a fixed external timescale
$\tau_3 \sim t_{\Omega}=1/\Omega$. 
This suggests an analogy between strong rotating turbulence 
and RMHD (i.e. MHD turbulence with a strong mean field). 
The simulations presented here show that the {\it weak limit} regime of 
$E_{k_\perp} \sim k_{\perp}^{-2}$ is obtained 
when $t_A/t_p \approx 1/2$. 
For larger values of this timescale ratio, the energy spectrum approaches
a Kolmogorov-like regime $E_{k_{\perp}} \sim k_{\perp}^{-5/3}$. 

For the small $t_A/t_p=0.1$ case the spectrum is 
$E_{k_{\perp}} \sim k_{\perp}^{-2.4}$ which would indicate that
the weak turbulence regime gave rise to a regime in which the
perpendicular cascade is suppressed even more.
In these numerical experiments we have kept 
the value of $S=\l0^2/(\eta\tA)$ fixed, while changing
the value of $\tA/\tp= (L_z U_p) / (\l0 \vA)$. If we consider that
all parameters are fixed, except for $U_p$, we can interpret
these runs as different evolutions of the system when varying the
intensity $U_p$ of the boundary driving. The result is that  
as the driving intensity is enhanced (i.e. larger $t_A/t_p$), the
spectrum becomes flatter and the regime goes through weak turbulence to  
a strong Kolmogorov-like state. As shown
by \Fig{3} (left panels), the kinetic energy spectrum is always
smaller than the magnetic energy spectrum 
and the ratio $E^u_k/E^b_k$
is smaller (especially at the low $k$ structures) when the forcing is weaker. 

The phenomenological argument expressed
by \Eqn{phenom} can be modified to address the very weak forcing
case. For that case, the energy at the low and intermediate $k$ values
is essentially magnetic $E_{k_{\perp}} \approx E_{k_{\perp}}^b$ (see \Fig{3}). 
For the non-linear time $\tau_{nl} (k) \sim (k u_k)^{-1}$, we replace the 
$k$-dependent velocity $u_k$ by a constant which we expect to be of the order of
the imposed boundary weak velocity $U_p$ (i.e. $\tau_{nl} \sim (k_{\perp} U_p)^{-1}$).
A scale invariant characteristic speed corresponds to a kinetic energy spectrum 
$E_{k_{\perp}}^u \sim k_{\perp}^{-1}$.
Again assuming $\tau_3 \sim t_A$ in \Eqn{phenom} we find, 
\begin{equation}
\epsilon \sim  t_A k_{\perp} E_{k_{\perp}}^b k_{\perp}^2 U_p^2
\end{equation} 
and a steep spectrum $E_{k_{\perp}}^b \sim k_{\perp}^{-3}$ is obtained. 
We should note that 
there are theoretical developments predicting spectra steeper than 
$k^{-2}$, due to nonlocal effects \cite{NgBhattacharjee97}, 
or due to high values of the correlation between magnetic and velocity fields
\cite{GrappinEA83}. The latter one corresponds to
$\langle u^2 \rangle \sim \langle b^2 \rangle$, 
a situation that would not be applicable for the present case.
The effects of the finite values of $S$ considered here are also expected
to be more important for very weak forcing.

\begin{figure}
   \caption[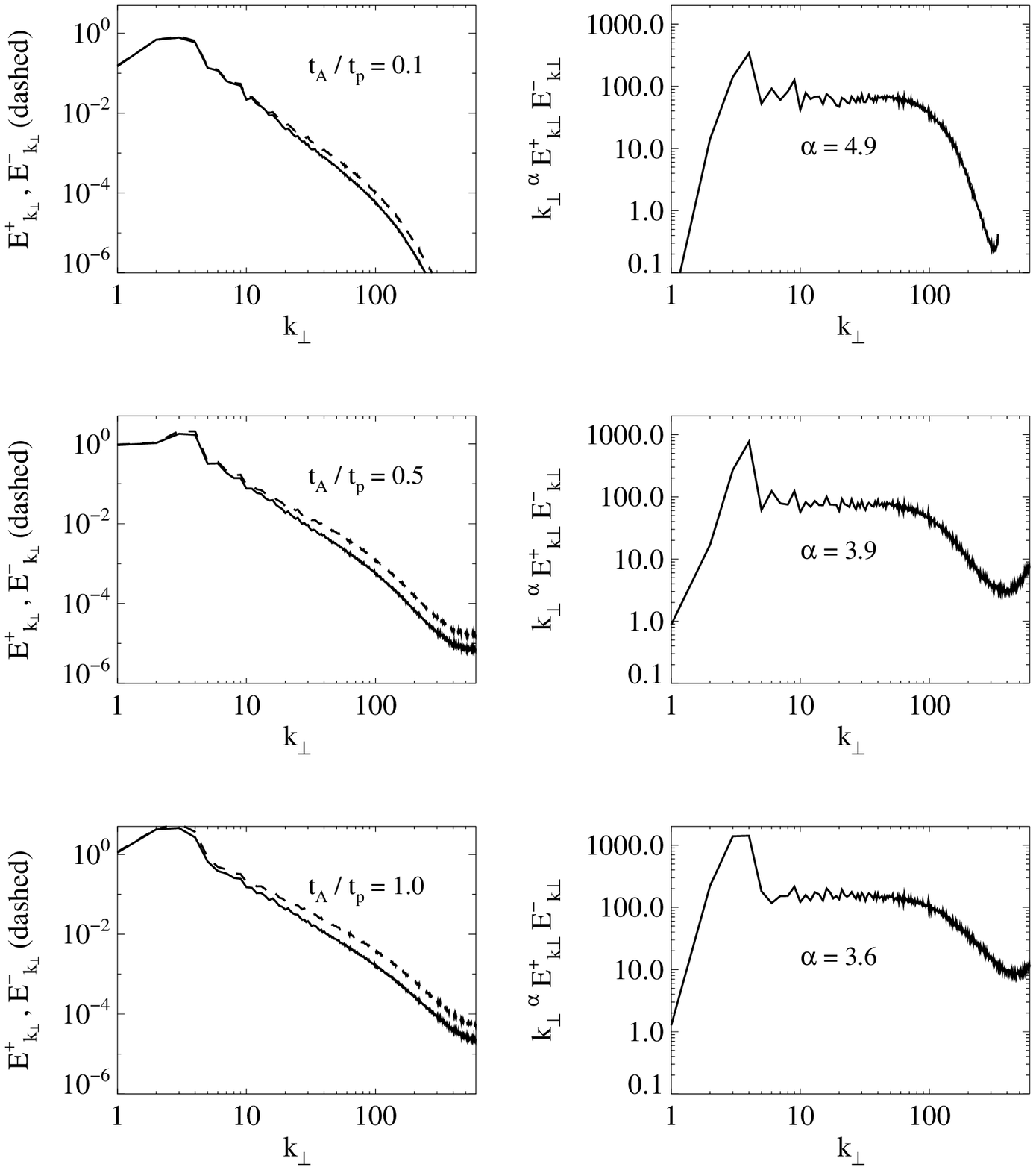]{
     Perpendicular energy spectra for the Elsasser fields ${\bf z}^{\pm}$ 
     from $S=2000$ runs and different 
     values of the ratio $t_A/t_p=0.1,~0.5,~1$. 
     On the left panels, $E^+_{k_{\perp}}$ and
     $E^-_{k_{\perp}}$ (dashed) spectra. 
     On the right, compensated
     spectra for the product $E^+_{k_{\perp}} E^-_{k_{\perp}}$ 
     and corresponding power index value $\alpha$.
    }
   \label{fig:5}
\end{figure}

Phenomenologies \cite{NgBhattacharjee97} and closures \cite{Galtier00}
are also usually presented in terms of the Elsasser fields 
${\bf z^{\pm}} = {\bf u} \pm {\bf b}$, so for completeness 
we computed the corresponding perpendicular spectra
\begin{equation}
E^{\pm}_{k_{\perp}} = \frac{1}{4} \int \sum_{k_{\perp} < k_x^2+k_y^2 <
                         k_{\perp} + \Delta k}
                         |{\bf z^{\pm}}(k_x,k_y,z)|^2 ~dz
\end{equation}
The results for the cases $t_A/t_p=0.1, 0.5, 1$ are shown in \Fig{5}.
The compensated spectra for the product $E^+_{k_{\perp}} E^-_{k_{\perp}}$
are shown on the right panels of \Fig{5}. Weak turbulence 
theory \cite{Galtier00} predicts an index $\alpha=4$, which for this 
particular problem corresponds again
approximately to the case $t_A/t_p= 1/2$. A higher power index 
$\alpha \sim 4.9$ is instead obtained for the very weak forcing
case $t_A/t_p=0.1$. 

\begin{figure}
   \caption[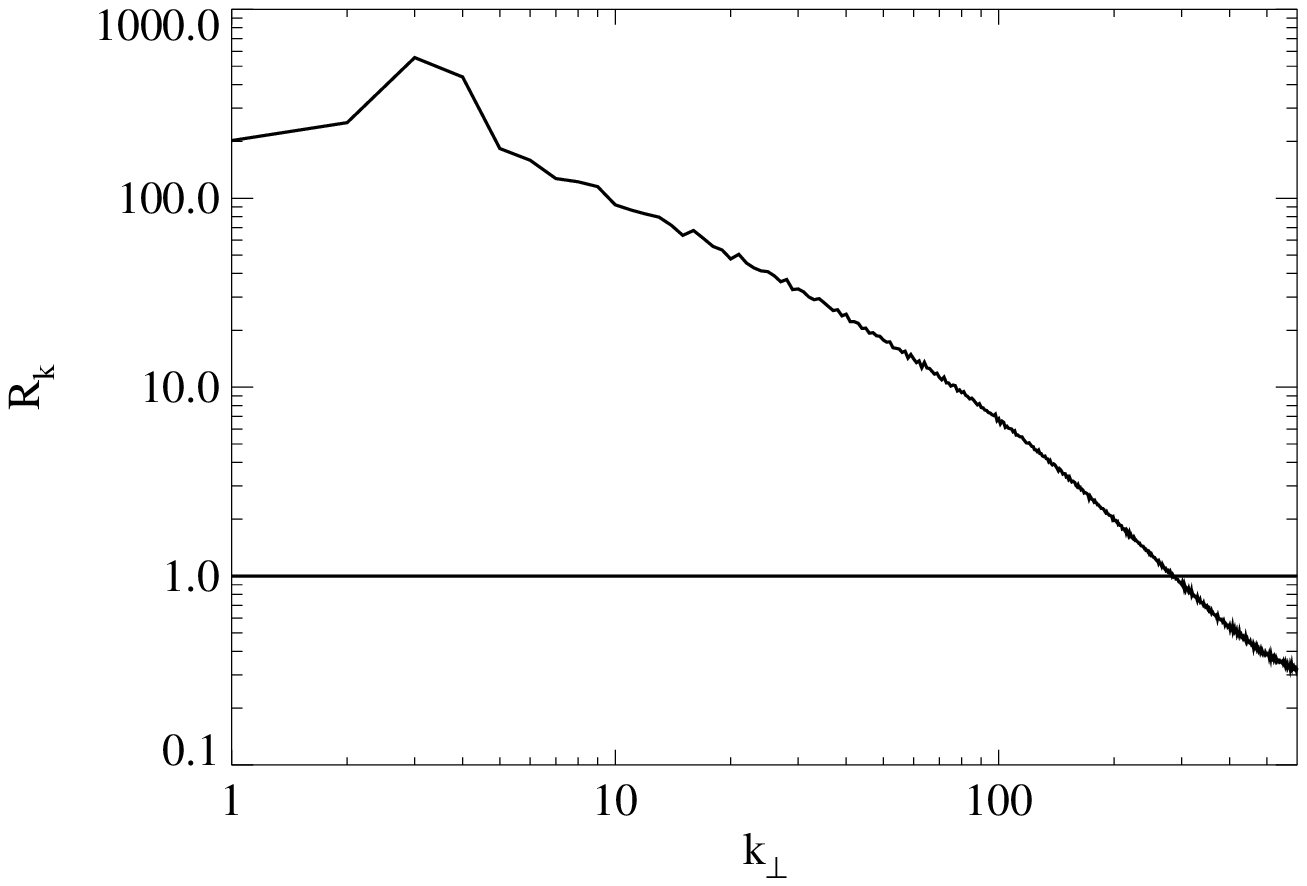]{
The $k$-dependent kinetic Reynolds number
$R_k = u_k/(k~\nu)=\sqrt{(E_k^u/k)}/\nu$
for a run with $S=2000,~t_A/t_p=1$.
    }
   \label{fig:6}
\end{figure}

\begin{figure}
  \caption[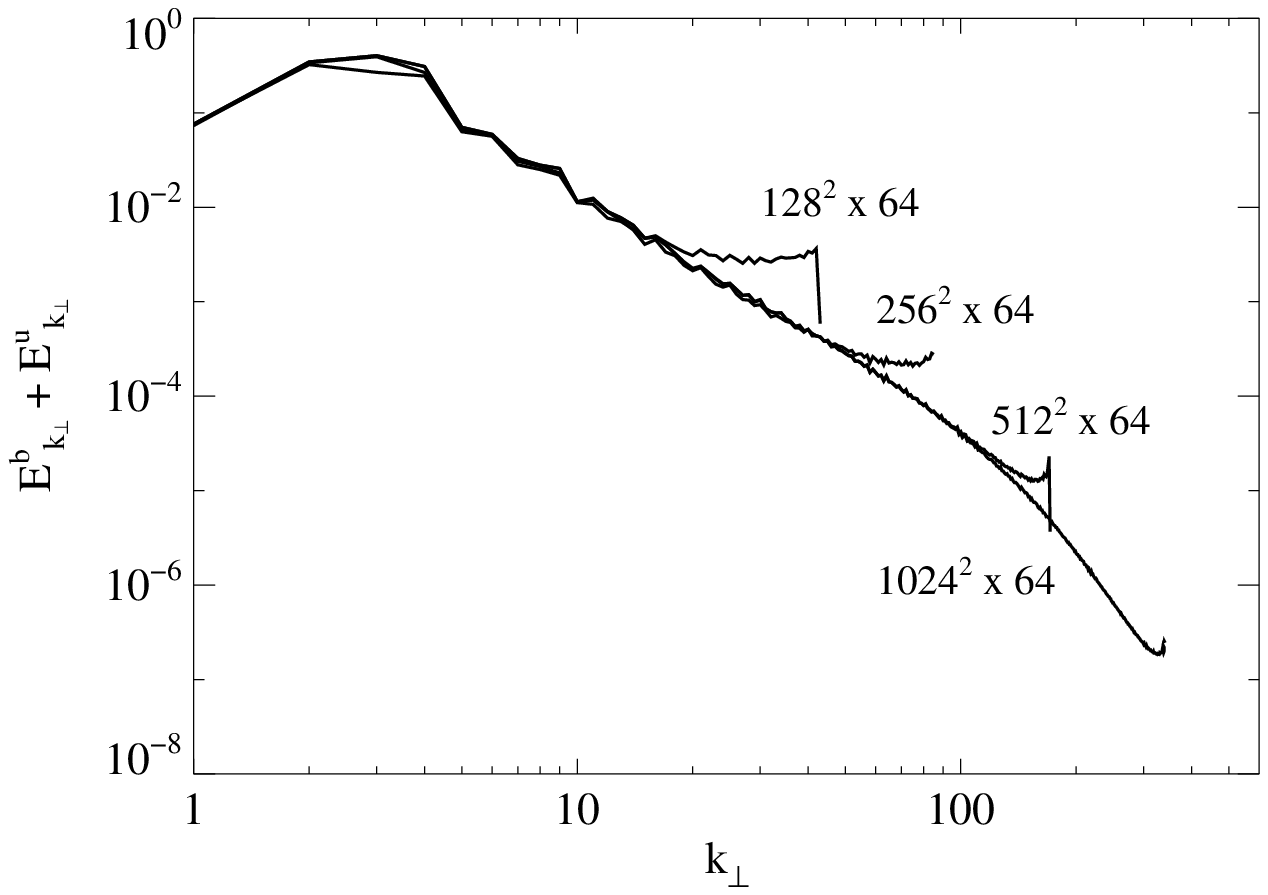]{Resolution study of the spectra for runs with
   $S=2000$, $t_A/t_p=0.1$
    }
   \label{fig:7}
\end{figure}

A rather technical but important issue when determining spectra is whether
the dissipation range is well resolved. We have considered here an
exact laplacian dissipation term in the dynamical equations (i.e. no
hyperviscosity is used) which makes harder to 
achieve higher spatial resolutions. 
An example from a simulation with $t_A/t_p=1$ is shown
in \Fig{6}. A $k$-dependent Reynolds number is computed as 
$R_k = u_k/(k~\nu)=\sqrt{(E_k^u/k)}/\nu$. The dissipative region in
$k$-space is characterized by
$R_k\sim 1$. The fact that $R_k$ reduces monotonically and falls below
unity at the largest values of $k$, is compelling evidence that
the dissipation range is well resolved in our simulations. 
Similar results are obtained for
the different values of $t_A/t_p$, although higher perpendicular resolution
is required as this value is increased. For $t_A/t_p=2$ the 
excessive computational requirement has forced us to consider a lower
value of $S=800$ to resolve the dissipation range.
Another direct resolution study is shown
in \Fig{7} where different perpendicular spatial resolutions have been adopted
for a simulation with $t_A/t_p=0.1$. What this figure shows is that
the spectrum is better resolved and extends to higher wavenumbers as the
resolution is increased. However, a very good picture of the lower $k$ part
of the spectra can still be obtained with the low resolution runs.

\section{Spatial structures}

\begin{figure}
   \caption[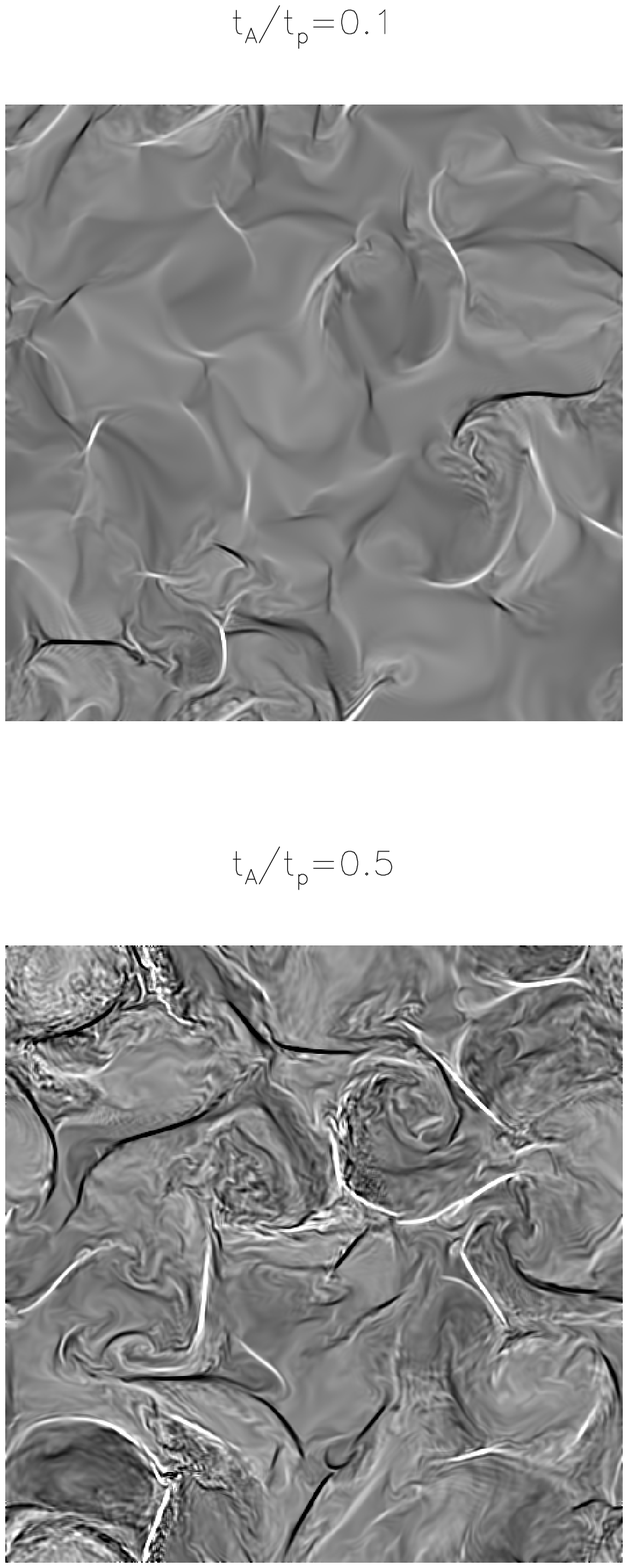]{
    Cross section of parallel current density for cases with 
    $t_A/ t_p =0.1$ and $t_A / t_p=0.5$.
    }
   \label{fig:8}
\end{figure}

\Fig{8} is a cross section of the simulation box (perpendicular
to the strong external magnetic field) showing
the spatial distribution of electric current density
(z-component) at $t=120 \tA$ for the cases $\tA/\tp=0.1$ (\Fig{8}a) 
and $\tA/\tp=0.5$ (\Fig{8}b). 
Intense positive z currents correspond to
white regions, while intense negative z current concentrations
are indicated in black. 
The current density structures extend in the z-direction (not shown) 
with almost no variation (as consistent with the RMHD approximation) 
in a current sheet-like form.
These structures are highly dynamic in the $x,y$ plane and evolve in time.
They indicate the presence of multiple reconnection events (where
the reconnecting magnetic field lies in the plane perpendicular
to the external field) occurring within the turbulent non-linear
dynamics.

The minimum thickness that current sheets can reach is determined 
by the magnetic diffusivity. In numerical simulations of turbulent regimes,
the magnetic diffusivity is usually made as small as possible,
in such a way that the sheet thickness (which is about the smallest
spatial feature expected in these simulations) is just marginally resolved.
The two cases presented in \Fig{8} share the same value of magnetic diffusivity
($S=2000$), and in fact it is possible to
find very narrow current 
sheets in both cases, close to the resolution limit. 
Both cases correspond to statistically steady regimes.
Nonetheless, 
the case corresponding to the less intense forcing
($\tA/\tp=0.1$ in \Fig{8}a), shows a comparatively fewer 
number of current sheets. Therefore, the number of current sheets
seems to be controlled by the intensity of the forcing. 
The widths of these current sheets on the other hand, 
are mostly determined by the 
dynamics. As a result, in both cases a wide range of 
current sheet widths can be observed.

\section{Conclusions}

We presented results of the turbulent dynamics of a bundle of magnetic flux ropes 
driven at their endpoints with steady convective motions. 
Two dimensionless parameters controlling
the response of the system are identified: the dissipation coefficient
and the ratio between the Alfven wave box crossing time to the forcing
timescale, $\tA / \tp$. As we showed in a previous work \cite{DmitrukGomez99},
the dissipation rate strongly depends on the value of this timescale ratio,
and is only weakly dependent on the dissipation coefficient.
It is found that the external driving generates a broadband 
perpendicular energy spectrum
with a slope which is determined by the timescale ratio $\tA / \tp$.
For small values of this ratio, a spectrum with slope $\alpha \approx -2.4$
is obtained. 
For $\tA / \tp \sim 0.5$ the spectrum is $ \sim k_{\perp}^{-2}$,
which would indicate a realization of the regime known as 
weak turbulence.
For higher values of $\tA / \tp$ the spectrum approaches the Kolmogorov
form $k_{\perp}^{-5/3}$. Both the weak turbulent and the Kolmogorov spectra
can be obtained from a common phenomenological framework based
on the assumed value of the time for decay of triple correlations.
The regime corresponding to spectral power laws
with slope -2.4 and lower for low values of $\tA / \tp$ should
be attributed to the relatively small amplitude flow at the boundaries.
The phenomenological framework can be modified to consider 
this weak velocity field case and a steep 
magnetic spectrum $\sim k_{\perp}^{-3}$ is obtained.
We also show the development of small scale spatial structures, in
the form of current sheets oriented along the mean field. 
The number of current sheets
seems to be controlled by the timescale ratio $\tA/ \tp$, while 
the widths of these current sheets on the other hand, 
are mostly determined by the 
dynamics. A wide range of current sheet widths can be observed.
The system presented here should be consider as an example of 
the effect of the boundaries on MHD turbulence. 
To our knowledge, there is a lack of
such studies in the literature 
and further investigation along these lines should be
extended to other MHD systems (3D MHD, compressible) as well.

\acknowledgements
Research supported by NASA NAG5-7164, NSF ATM-0105254 and ATM-9977692.
DOG is a researcher of CONICET (Argentina) and acknowledges 
support from grants UBACYT X209 
(University of Buenos Aires, Argentina) and PICT 03-9483 (ANPCyT, Argentina).



\end{document}